\documentclass[journal]{IEEEtran}
\usepackage[square,sectionbib]{natbib}
\usepackage{amssymb}
\usepackage{amsmath}
\usepackage{graphicx}
\usepackage{amsfonts}
\newtheorem{definition}{Definition}
\newtheorem{lemma}{Lemma}
\newtheorem{theorem}{Theorem}

\ifCLASSINFOpdf
\else
\fi
\begin{document}
%
\title{Facticity as the amount of self-descriptive information in a data set}
%
%
%

\author{Pieter~Adriaans

\thanks{P.W. Adriaans is with the ILLC and IvI, Department of Computer Science,
University of Amsterdam, Science Park 904,
1098 XH  Amsterdam, The
Netherlands. emailL P.W.Adriaans@uva.nl}
\thanks{Manuscript received March, 2012}
}

\markboth{arXiv}%
{Adriaans: Facticity as the amount of self-descriptive information in a data set}
%



\maketitle

\begin{abstract}
Using the theory of Kolmogorov complexity the notion of \emph{facticity} $\varphi(x)$ of a string is defined as the amount of self-descriptive information it contains. It is proved that (under reasonable assumptions: the existence of an empty machine and the availability of a faithful index) facticity is definite, i.e. random strings have facticity $0$ and for compressible strings  $0 < \varphi(x) < 1/2 |x| + O(1)$. Consequently facticity measures the tension in a data set between structural and ad-hoc information objectively. For binary strings there is a so-called facticity threshold that is dependent on their entropy. Strings with facticty above this threshold have no optimal stochastic model and are essentially computational. The shape of the facticty versus entropy plot coincides with the well-known sawtooth curves observed in complex systems. The notion of factic processes is discussed. This approach overcomes problems with earlier proposals to use two-part code to define the meaningfulness or usefulness of a data set.
\end{abstract}


%
\IEEEpeerreviewmaketitle

\section{Introduction}

\IEEEPARstart{A}{ll} known formal measures of information (Shannon \cite{CT2006}, Kolmogorov \cite{LiVi08}, Fisher \cite{FIS25}) assign the highest information value to data sets with maximum entropy. This implies that a television broadcast with pure noise is the most information-rich program we can watch. This obviously does not cover our intuitions about what useful information is. In the past decennia there have been a number of competing proposals to define a formal unit of measurement of meaningful or useful information.
\begin{itemize}
  \item Esthetic Measure (Birkhoff, Bense \cite{Ben65}, \cite{Bir50}, \cite{Scha93}).
  \item Sophistication (Koppel, \cite{Kop87}, \cite{Ant03}, \cite{Ant06})
  \item Logical Depth (Bennet, \cite{Ben88})
  \item Statistical complexity (Crutchfield, Young , \cite{CJ89}, \cite{CJ90}, \cite{CR94})
  \item Effective complexity (Gell-Mann and Lloyd, \cite{GL2003})
  \item Meaningful Information (Vit\'anyi, \cite{Vit06})
  \item Self-dissimilarity (Wolpert and McReady, \cite{WM2007})
  \item Computational Depth (Antunes et al., \cite{Ant06})
\end{itemize}
Three intuitions dominate the research. A string is 'interesting' when ...:
\begin{itemize}
  \item A certain amount of computation is involved in its creation (Sophistication, Computational Depth).
  \item It has internal phase transitions (self-dissimilarity).
  \item There is a balance between the model-code and the data-code under two part code optimization (effective complexity).
\end{itemize}
Such models penalize both maximal entropy and low information content, but the exact relationship between these intuitions is unclear.  Several authors have suggested to use Rissanen's notion of Minimum Description Length (MDL) (\cite{RIS78}, \cite{RIS89}) and the theory of Kolmogorov complexity (\cite{Vit06}) as building blocks for a theory of meaningful information (\cite{LiVi08}, \cite{GL2003}). This idea is already implied in Kolmogorov's structure function \cite{VV02}. There are fundamental problems with this approach (\cite{MA2003}). One quality a theory of facticity\footnote{The term 'facticity' is derived from Heidegger, and denotes the unexplainable 'givennes' of reality. The roots of this notion are theological: 'factum est': it has been made. Conf. "All things were made by him; and without him was not any thing made that was made": "Omnia per ipsum facta sunt, et sine ipso factum est nihil, quod factum est" (Gospel of St. John: 1,1,3). I think the term is appropriate because the theory of facticity is as close as we will ever come to a pure mathematical theory of creation and creativity.} ought to have is that is should give us a guarantee that it is an objective measure. It should be \emph{definite}: 1) It should extract all model information from a data set but 2) not more. To my knowledge, none of the proposals for meaningful information made so far in the literature have been proved to be definite.

\section{Definitions}
\subsection{Kolmogorov Complexity}
We will follow the standard textbook of Hopcroft, Motwani and Ullman for the basic definition of a \emph{Turing machine} (TM)(\cite{HU2001}). $\mathcal{T}$ will be the set of all possible descriptions according to this formalism.
\begin{definition}[Self-delimiting Code]\label{SelfDel}
Let $x$ be a binary string, the self-delimiting code for $x$ is defined as the binary string $\overline{x}=0^{\log c}1cx$ where $c = |x|$ is (a binary representation of) the length of $x$.  We have $|\overline{x}| = c +  2\log c + 1$. The self-delimiting code for the empty string is $\overline{\varepsilon}=1$.
\end{definition}
An example: the self-delimiting code for "11001110" is the concatenation of "000","1","100" and"11001110". Note that self-delimiting code in this sense is prefix-free: strings of different length get a different prefix. Throughout this paper we will assume a reference universal Turing machine $\overline{U}$ with prefix-free indexes has been chosen. If $\overline{U}(\overline{\imath}p)=x$, then $\overline{\imath}$ is the prefix-free index of a Turing machine $T_i$ to be emulated by $\overline{U}$ and $p$ is a input string for $T_i$. Note that the input string $p$ is not prefix-free, a feature that is essential for the results in this paper. In short: $\overline{U}(\overline{\imath}p)=T_i(p)=x$  All definitions in this paper refer to the preselected machine $\overline{U}$ which we will refer to as $U$. Without loss of generality we will suppose that there is a minimal Turing machine $T_0$ that simply is empty and does not compute anything. The index of $T_0$ is the empty string $\varepsilon$.
\begin{lemma}[Swap-machine]\label{Swap}
 There exists a special machine with index $s$ that simply swaps the index and the input, i.e. For all $p$ if $U(\overline{\imath}p)= x$ then: \[U(\overline{s}\ \overline{p}i)=U(\overline{\imath}p)= x\]
\end{lemma}
Proof: This follows from the fact that $U$ is universal.
$\Box$

\begin{definition}\label{OPCODE}
Let $x$ be a binary string and let $U$ be a universal Turing machine. The \emph{optimal code} for $x$ is the shortest code that generates $x$ on $U$:
\[x^{\ast}= \min_{i}\{i:U(i)=x\}\]
\end{definition}
The length of the optimal code defines the classical Kolmogorov complexity:
\begin{definition}\label{KOLCLAS}
The \emph{Classical Kolmogorov complexity} of a binary string is: $C(x)= \min\{|i|:U(i)=x\}$
\end{definition}
\begin{definition}\label{KOL}
The \emph{prefix-free Kolmogorov complexity} of a binary string is: $K(x|y)= \min\{|\overline{\imath}|:U(\overline{\imath}y)=x\}$
\end{definition}
We define:
\begin{definition}
$K(x)= K(x|\varepsilon)$
\end{definition}
This is in fact a one-part code optimization variant $K_1$ of Kolmogorov complexity that forces all complexity of the information to be stored in the index of the Turing machine. It is useful to distinguish a two-part code optimization variant:
\begin{definition}\label{KOL2}
$K_2(x)= \min_{i,p}\{|\overline{\imath}|+|p|:U(\overline{\imath}p)=x\}$
\end{definition}
This version balances the information over an index $i$ and a program $p$ for $x$. Here $\overline{\imath}$ is the self-delimiting code of an index and $U$ is a universal Turing machine that runs program $p$ after interpreting $\overline{\imath}$ and $\varepsilon$ is the empty string.The reason to use $\overline{\imath}$ lies in the fact that it allows us to separate the concatenation $\overline{\imath}p$ into its constituent parts, $i$ and $p$. Here $i$ can be seen as capturing the \emph{regular} (structural \cite{VV02}, meaningful \cite{Adriaans2009}, model \cite{Grunwald:2007:1}, effective \cite{GL2003}) part of the string $x$, where $p$ describes the \emph{irregular} part.

The following lemmas show that two-part code is really more expressive than its one-part variants.
\begin{lemma}\label{KK2}
For all $x$ we have $K(x) \geq K_2(x)$.
\end{lemma}
Proof: Suppose $K_2(x) > K(x)$. We have $K_2(x)= \min_{i,p}\{|\overline{\imath}|+|p|:U(\overline{\imath}p)=x\}$ , $K(x)= \min_{j,\varepsilon}\{|\overline{\jmath}|:U(\overline{\jmath}\varepsilon)=x\}$ i.e. $|\overline{\imath}|+|p| > |\overline{\jmath}|$, but then $K_2(x)= \min_{j,\varepsilon}\{|\overline{\jmath}| + 0:U(\overline{\jmath}\varepsilon)=x\} = K(x)$. $\Box$

\begin{lemma}\label{CK2}
For all $x$ we have $C(x) = K_2(x)$.
\end{lemma}
Proof: We have $K_2(x)= \min_{i,p}\{|\overline{\imath}|+|p|:U(\overline{\imath}p)=x\}$, $C(x)= \min_{q}\{|q|:U(q)=x\}$. Note that both $C$ and $K$ are defined with respect to the same prefix-free Turing machine $U$, which implies that  $\overline{\imath}p = q$. $\Box$

The elegance of the introduction of an empty machine is illustrated by:
\begin{lemma}\label{EmptyRandom}
For any random string $x$ we have $K_2(x)= |x| + 1$.
\end{lemma}
Proof: Suppose $x$ is random. In this case it cannot be compressed and the empty machine $T_0$ with index $\varepsilon$ is the best model of length $0$. We have $U(\overline{\varepsilon}x)=x$ and thus $K_2(x)= |x| + 1$. $\Box$

There exists a so-called universal distribution $m$ with $m(x) = 2^{- (K(x) +O(1))}$. This distribution dominates any recursive distribution by a multiplicative constant \cite{LiVi08}.

\subsection{Information theory}
We will follow the standard textbook of Cover and Thomas for the basic definitions of  \emph{Information theory} (\cite{CT2006}).
\begin{definition}
A \emph{binary source code} $C$ for a random variable $X$ is a mapping from $\mathcal{X}$, the range of $X$, to $\{0,1\}^{\ast}$. Let $C(x)$ denote the codeword corresponding to $x$ and let $l(x)$ denote the length of $C(x)$. The expected length $L(C)$ of a source code $C(x)$ with probability mass function $p(x)$ is given by $L(C)  = \sum_{x in \mathcal{X}} p(x)l(x)$.
\end{definition}
The following lemma (\cite{CT2006}, lemma 5.8.1) is important:
\begin{lemma}\label{existence-of-optimal-compression}
For any distribution, there exists an optimal instantaneous code (with minimum expected length) that satisfies the following properties:
\begin{itemize}
\item The lengths are ordered inversely with the probabilities (i.e. if $p_j > p_k$ then $l_j \leq l_k$).
\item The two longest codewords have the same length.
\item Two of the longest words differ only in the last bit and correspond to the two least likely symbols.
\end{itemize}
\end{lemma}
For a system of messages $S$ with $\{s_i : s_1, s_2,...,s_n\}$, the Shannon entropy is defined as
\[H(S)=- \sum_{i=1}^{n} p(s_i) log_2p(s_i)\] Here $p(s_i)$ is the probability of message $s_i$. The following concept is useful:
\begin{definition}[Inverse Entropy]\label{Inverse-entropy}
The entropy for binary strings based on a system of messages $S$ with probability $p$ is $H(p)= - p \log p - (1-p) \log (1-p)$. The inverse entropy on the interval $s \in [1/2, 1]$ can be estimated using  \[H'^{-1}(s)= - \frac{s}{W(s)} - \frac{1-s}{W(1-s)}\] here $W(x)$ is the product log function. Because of the symmetry of the entropy versus probability plot, we can use the value $p = H^{-1}(s)=1-H'^{-1}(s)$ that is defined on the interval $[0,1/2]$ to find the probability $p$ associated with a certain entropy.
\end{definition}

\begin{definition}\label{typical}
A \emph{stochastic binary string} is a binary string generated by a system of messages $S=\{0,1\}$ with a certain entropy $H(S) \leq 1$.
\end{definition}
Stochastic binary strings define the connection between Shannon information and Kolmogorov complexity:
\begin{lemma}\label{limit}
For a stochastic binary string of length $k$ we have in the limit $K(x)=H(x)=kH(S)$.
\end{lemma}
Proof:
We use a result from \cite{CT2006} (Theorem 5.4.2.). Since $S=\{0,1\}$ is a stationary stochastic source the expected code length per symbol is $H(S)$, which gives $kH(S)$ as optimal compression length. Since $x$ is stochastic we have $H(x) + |\overline{\imath}|=K(x)$, where $i$ is the index of a program that prints $x$ on the basis of an optimal code. Since  $|\overline{\imath}|$ is constant in the limit we have $H(x)=K(x)$. $\Box$

An useful lemma is:
\begin{lemma}\label{compressible}
Almost all compressible binary strings are stochastic.
\end{lemma}
Proof: Consider the set of binary strings $X^n_k$ of length $n$ with $k$ zeros. These strings can be enumerated using an index of length $\log_2 {n \choose k}$. A vanishing fraction of these indexes is itself compressible. A program of $p$ constant length transforms the indexes in to the original strings giving $K(x) \leq \log_2 {n \choose k} + |p|$ for elements of $X^n_k$. The majority of strings in $X^n_k$ will be typical: $K(x) = \log_2 {n \choose k} + |p|$. In the limit the contribution of $|p|$ vanishes. $\Box$\\

These limit results are rather rough and can be refined once we have formulated a definite measure for the model information in a string.

\section{Facticity}
A problem that has hampered all proposals to define meaningful information in terms of two-part codes is the fact that recursive indexes for Turing machines never reflect the true model information . This is known as the nickname-problem (\cite{GL2003}, \cite{Foley2010}, \cite{MA2003}). The following lemma's show that we can in almost all cases assume that the index function we use is faithful.
\begin{definition}[Faithful index function]\label{faithful}
Let $I$ be an index function that gives an index for all elements of a set of descriptions of all Turing machines $\mathcal{T}$ according to some formalism (say the one introduced in Hopcroft and Ullmann). $I$ is \emph{faithful} to $\mathcal{T}$ iff we have for all $\forall(t \in \mathcal{T}):C(t) \leq |I(t)| \leq C(t)+O(1)$, i.e. the length of the index reflects the Kolmogorov complexity of the machine within a constant.
\end{definition}
\begin{lemma}[Existence of faithful index]\label{Existence}
For every universal Turing machine $U$ a faithful index set $I$ exists, but it is not recursive.
\end{lemma}
Proof: Take $I'$ to be the set of shortest programs for elements of $\mathcal{T}$ that generate descriptions (i.e. $U(i')=t$) and let $q$ be a program that interprets elements of $\overline{q}i \in I$ such that $U(\overline{q}\ \overline{\imath}p)=T_i(p)$, i.e. $q$ reads in $i$, converts it to a description $t$ and emulates $t$ on $U$ in order to process $p$. For all elements if $\overline{q}i \in  I$ we have $C(t) \leq |\overline{q}i| \leq C(t)+O(1)$. Since $C(t)$ is not computable such an index can never be recursive.
$\Box$

In the following paragraphs we will assume that the index functions are faithful. The following lemma shows that this is in almost all cases a good assumption:
\begin{lemma}[Recursive definitions are almost always faithful]
Let $q$ be a optimal program that generates a unique string $x_n$ on $U$ for each natural number $n$, then for almost all $n$ the program $\overline{q}x_n$ is a faithful index for $U(\overline{q}x_n)$.
\end{lemma}
Proof: The faithfulness condition is $C(U(\overline{q}x_n)) \leq |\overline{q}x_n| \leq C(U(\overline{q}x_n))+O(1)$. The first inequality holds by definition. Suppose that $|\overline{q}x_n| > C(U(\overline{q}x_n))+O(1)$. In this case, since $q$ is optimal, $U(\overline{q}x_n)$ is compressible below  $|\overline{q}x_n| - O(1)$. Since the density of compressible strings in the limit is zero, this event is extremely rare. $\Box$

We cannot design recursive index functions that are systematically non-faithful. The facticity of a string $x$ is defined as:

\begin{definition}[Facticity]\label{facticity}
\[\varphi_U(x)= \min \{|i|: \exists(p)(|\overline{\imath}p|= K_2(x) \ \& \ U(\overline{\imath}p)=x)\}\]
\end{definition}

i.e. the length of the shortest model code of all optimal models under two-part code optimization. Note that the additional code length necessary to make the model prefix-free is not taken in to account in the definition of facticity. This seems reasonable since this code is not part of the content of the model per se. Intuitively $\varphi(x)$ is a measure of how 'interesting' or 'useful' the string $x$ is. Note that there might be different models that produce the same facticity. We generally feel that random data-sets do not contain much meaningful information. This behavior of facticity is illustrated by the following theorem:
\begin{lemma}\label{Minimality1}
For any random string $x$ we have $\varphi(x)=0$.
\end{lemma}
Proof: This follows from lemma \ref{EmptyRandom}.
Suppose $x$ is random. We have $U(\overline{\varepsilon}x)=x$ and thus $\varphi(x)= |\varepsilon|=0$. $\Box$

Note that the reverse selection $p=0$ and $i=x$ would penalize the total code length with a factor $O(\log x)$, so this choice is never made. This lemma shnows that, in terms of facticity, random strings are not meaningful. The reverse of this lemma is:
\begin{lemma}\label{Model}
All compressible strings have a non-empty model: if $K_2(x) < |x|$ then $\varphi(x) > 0$.
\end{lemma}
Proof: Suppose $x$ is compressible and that $\varphi(x) = 0$. We have
$K_2(x)= \min_{i,p}\{|\overline{\varepsilon}|+|p|:U(\overline{\varepsilon}p)=x$. Since $p$ is the shortest possible code it is random, moreover, after processing the first part of the input $\overline{\varepsilon}$  the machine $U$ will not start a new computation. Consequently  $x = p$. This contradicts the fact that $x$ is compressible. $\Box$

On the other hand the maximal amount of meaningful information in a string is limited to half its length plus a constant:

\begin{lemma}\label{Minimality2}
For any string $x$ we have $\varphi(x) \leq 1/2 K_2(x) + |\overline{s}|$ where $s$ is the swap-machine.
\end{lemma}
Proof: $K_2(x)= \min_{i,p}\{|\overline{\imath}|+|p|:U(\overline{\imath}p)=x\}$. Suppose $\varphi(x) > 1/2 K_2(x) + |\overline{s}|$ then there are $p$ and $i$ such that $U(\overline{s}\overline{p}i)=U(\overline{\imath}p)= x$, but then since $|i| \gg |p|$ we have $|\overline{s}\overline{p}i| < |\overline{\imath}p|  + |\overline{s}| = K_2(x) + |\overline{s}|$ and consequently $|\overline{s}\overline{p}i| < K_2(x)$ which contradicts the fact that $K_2$ gives the length of the shortest code.  $\Box$

We now have to prove that facticity really covers the concept of the exact amount of meaning in a string. This is summarized in the main theorem of this paper:

\begin{theorem}[Facticity is definite]\label{DEF}
 If a string $x$ is compressible then $0 < \varphi(x) < 1/2 |x| + O(1)$. If $x$ is random then $\varphi(x)=0$.
\end{theorem}
Proof: this follows from lemma's \ref{Minimality1}, \ref{Model} and \ref{Minimality2}.  $\Box$

This shows that facticity is actually the result of a balance between certain tensions in the data set: If the model code becomes too short, we loose to much of the computational power of our universal Turing machine, if it becomes too long, the price for separating the program code from the data becomes too high. These intuitions seem reasonable. We can now give the following informal definition:
      \begin{definition}
      The \emph{facticity} $\varphi(x)$ of a string $x$ is the amount of self-descriptive information $x$ contains.
      \end{definition}

\subsection{Some results}
In this paragraph I investigate the relation between complexity $K_2(x)$ and factictiy $\varphi(x)$.

 \begin{definition}
Given a string $x$ with prefix-free Kolmogorov complexity $K_2(x)$ and facticity $\varphi(x)$ I give the following definitions:
\begin{itemize}
\item The \emph{randomness deficiency} of $x$ is $\delta(x) = |x| - K_2(x)$.
\item A string is \emph{non-stochastic} when $\varphi(x)=K_2(x)$.
\item A string is \emph{mixed} when  $0 < \varphi(x) < K_2(x)$.
\item The \emph{residual entropy} of a string is: $\rho(x)= K_2(x) - (\varphi(x) + 2 \log \varphi(x) +1)$.
\end{itemize}
\end{definition}

The distinction between stochastic and non stochastic strings is important. The following lemma describes the behavior of the stochastic strings.

\begin{theorem}\label{cut-off-point}
Binary stochastic strings have, with probability $1-\epsilon$, small optimal models of size $\log |x| + \log k + c$, where \[\epsilon = (1-(H^{-1}(s))^{k})^{\frac{|x|}{k}}\]
\end{theorem}
Proof: According to lemma \ref{existence-of-optimal-compression} there is always an optimal code for which 1) the two longest codewords have the same length 2) two of the longest words differ only in the last bit and correspond to the two least likely symbols. For a binary stochastic string the probability of the least likely blocks of length $k$ will be determined by their density of either ones or zeros. Thus, for a stochastic string that is long enough, the only relevant model parameters are an integer of length $\log |x|$ giving the total number of ones in the string and $k$ giving the size of the maximal block length. From these two parameters we can estimate the density. A program of constant length $c$ gives an optimal code. This approach collapses when sampling the strings with block-size $k$ when some random longest word blocks do not occur in $x$ and the optimal code has to be adapted.  $H^{-1}(s)$ estimates the smallest probability of either zeros or ones in the string. $(H^{-1}(s))^{k}$ is the lowest probability for a block of length $k$. The probability that this block is not selected is $1-(H^{-1}(s))^{k}$. The probability that this block is not selected in $\frac{|x|}{k}$ trials is $\epsilon = (1-(H^{-1}(s))^{k})^{\frac{|x|}{k}}$. This is the probability that the string does not have an optimal model in the sense of lemma \ref{existence-of-optimal-compression}. $\Box$\\

Note that in the limit $K_2(x)$ approaches $|x|H(S)$. When $H(S)=1=s$ we have $H^{-1}(s)=1/2$, i.e. the probability that strings with high entropy have no small models goes to zero in the limit. When $H(S)=1/2=s$ we have $H^{-1}(s)\approx 0.1$. The probability that a stochastic string has a small model increases exponentially with length of the string, and for strings of fixed length polynomially with the complexity of the string. In the limit (long strings with small block-size) the cut-off will be sharp and act like a sort of percolation threshold. Above a certain complexity the size of optimal models for random strings will decrease rapidly. A defining family of curves is what one could call the:
\begin{definition}[Collapse probability]\label{CP}
For binary stochastic strings we have: \[\Phi(H(S)) = \Phi(s) = (1-(H^{-1}(s))^{k})^{2^k}\]
where $k$ is the block size and the length of $x$ is $|x|=k2^k$.
\end{definition}
 The collapse probabilities, that are instantiations of theorem \ref{cut-off-point}, specify the probability that a string of length $k2^k$ with a certain entropy between zero and one does not have a short optimal model when sampled with a block size $k$. The definition is derived from the so-called Coupon collectors problem that specifies the number of trials a collector of $n$ coupons has to make to collect all $n$ coupons as $n\log(n)$.


In general one could say that the factors $H(S)=s$ and $H^{-1}(s)$ specify a balance entropy and inverse entropy, that $2^k$ specifies the size of a space and $k$ a sampling granularity. The Collapse probabilities define the facticity curves directly as the following theorem shows:
\begin{theorem}[Facticity threshold] \label{optimal-model-size}
The optimal model size for a stochastic binary string $x$ of length $k2^k$ and entropy $H(S)$ is:
\[\varphi(x) \leq \log {2^k \choose \ulcorner2^k\Phi(H(S))\urcorner} + k + \log k + c\]
\end{theorem}
Proof: A term of size $\log k$ specifies the string length $k2^k$ and the block size $k$. A term of length $k$ specifies the density of zeros. This allows us to estimate $H(S)$. From this we calculate an optimal coding for a string with $H(S)$. The term $\Phi(H(S))$ gives us the probability that some of the $2^k$ possible blocks do not occur in $x$, giving $2^k\Phi(H(S))$ as the size of the set of blocks that do occur in $x$. An optimal index for this set has complexity \[\log {2^k \choose \ulcorner2^k\Phi(H(S))\urcorner}\]A program of length $c$ computes an optimal model on the basis of these data. $\Box$\\

\begin{lemma}
The maximal facticity of stochastic strings of length $n$ is $O(n/\log(n))$.
\end{lemma}
Proof: Direct consequence of theorem \ref{optimal-model-size}. Note that the maximal complexity of $\varphi(x) = 2^k  + k + \log k + c$ is reached for the value $\Phi(H(S))=1/2$. Take $k2^k = n$. $\Box$\\

Another important insight is that the density of a facticity plot is completely determined by the sampling probability distribution. If we use some simple high entropy process for the generation of examples all our strings will have small models and high entropy. The interesting plots are generated by computational procedures:

\begin{lemma}\label{DENS}
A sample taken under the universal distribution $m$ from the total set of binary strings of length $k$ will have uniform distribution over the complexity interval $[K_2(x)= 0,K_2(x)= k]$.
\end{lemma}
Proof: Direct consequence of Levin's coding theorem: $m(x) = 2^{- (K(x) +O(1))}$. The exponential decay of $m$ is up to a constant factor equal to the increase of density of binary strings with $K(x)$, so the exponential decay in probability is balanced by an exponential increase in density.   $\Box$\\

Note that the universal distribution dominates any recursive distribution up to a multiplicative constant.
Remember that the structure of a model is  $K_2(x) =  (\varphi(x) + 2 \log \varphi(x) +1) + \rho(x)$. Theorem \ref{optimal-model-size} gives an optimal model size for $x$, but in general this size will not be reached because of the penalty of size $2 \log \varphi(x) +1$. The search algorithm behind $K_2$ will settle for an optimal exchange between bits stored in the residual entropy and the facticity. Purely non-stochastic strings are found on the line $\varphi(x)=K_2(x)$. Their density diminishes exponentially with growing $K_2(x)$. Anthunes and Fortnow have proved the existence of so-called absolutely non-stochastic strings that encode the halting set for all smaller strings \cite{Ant03}. So, close to the upper bound $\varphi(x)= |x|/2$ there are still non-stochastic strings, although we will with very high probability never sample them. Note that we can transform every non-stochastic string to a mixed string by simply flipping some bits and adding a list of locations of the flipped bits. This representation is very inefficient. There will be a band of mixed optimal models close to the non-stochastic models.

\begin{figure}[!t]
\centering
\includegraphics[width=3in]{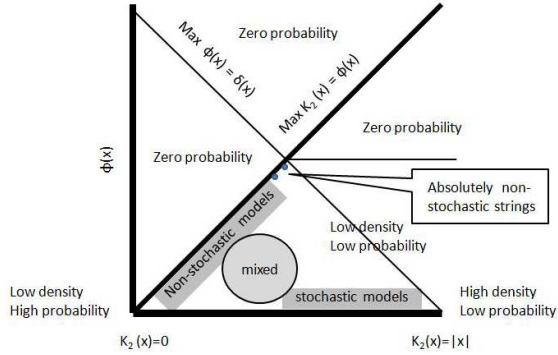}
\caption{Facticity density in the complexity space $[0,K_2(x) = |x|]$. The density estimates are compatible with the so-called saw-tooth curves observed in various data-sets.}
\label{Overview}
\end{figure}

We can give a taxonomy of strings under $K_2$ compression (See figure \ref{Overview}):
\begin{enumerate}
\item Strings below the facticity threshold (c.f. theorem \ref{optimal-model-size}) are either:
    \begin{enumerate}
        \item \emph{Non-stochastic}, i.e. the facticty is 'close' to the Kolmogorov complexity $\varphi(x) \approx K_2(x)$
        \item \emph{Purely stochastic}, i.e. the density of the strings is a sufficient model (see theorem \ref{cut-off-point}): $\varphi(x) = \log |x| + \log k + c$
        \item \emph{Stochastic}, i.e. they are under-sampled in the sense of theorem \ref{optimal-model-size}.
    \end{enumerate}
\item Strings above the stochasticity threshold (c.f. theorem \ref{optimal-model-size}) are in principle non-stochastic. They are either:
     \begin{enumerate}
        \item \emph{absolutely non-stochastic}, i.e. non-computable (c.f. \cite{Ant03}): $\varphi(x) \approx |x|/2$.
        \item \emph{Computable}, i.e. created by a non-deterministic process: $\varphi(x) \ll |x|/2$.
    \end{enumerate}
\end{enumerate}

The following lemma is interesting:
\begin{lemma}[Existence of a saturation point]\label{max-model}
 For any non-stochastic string $x$ with $\varphi(x)=K_2(x)$ we have \[\varphi(x) \leq 2^{|u|+ 2 \log |u| + 1}\] where $u$ is the optimal index of the smallest universal Turing machine $U$.
\end{lemma}
 Proof: $x$ is non-stochastic: $\varphi(x)=K_2(x) $, there are no data.  Note that $|u|+ 2 log |u| + 1$ is the length of the self delimiting code for $U$, as soon as the model becomes longer than $\varphi(x) = 2^{|u|+ 2 log |u| + 1}$ it is more efficient to prefix $U$ to the code as a general computational model and interpret the model as data.$\Box$\\

 This proof is related to the proof of lemma \ref{Minimality2}. An immediate consequence of this is that the maximal number of meaningful models that is available for pure computational structures is limited by: \[2^{2^{1 + u + \log u + \log\log u + \log\log\log u +....}}\] where $u$ is the length of the optimal index of the smallest universal Turing machine that exists and $u + \log u + \log\log u + \log\log\log u +....$ is the theoretical optimal length of its prefix-free index. Longer models will automatically be reinterpreted as data without structure. Only up to a certain limit there are non-stochastic objects, but there are \emph{mixed} strings with longer model information.

\subsection{Approximating facticity}
For approximation of facticity we have theoretically the same limits as were proved in \cite{Power09}: i.e. the facticity is in principle non-approximable in finite time. We may compress the data set, but there is no guarantee that the randomness deficiency of the model is also improved. In practice however, any learning or compression algorithm that allows us to estimate the balance between the data and the model code with reasonable accuracy can be used in real world data. Such algorithms include decision tree induction, nearest neighbor search, neural networks, grammar induction algorithms, standard compression algorithms etc. (\cite{Adriaans2009}, \cite{Adri06}, \cite{cilibrasi-clustering}).  The philosophical question why learning algorithms that in principle are faulty work reasonably well on real life data is still a matter of debate (\cite{Adriaans2009}, \cite{Grunwald:2007:1}).

Sawtooth plots are well-known in the literature and have been observed empirically in many different environments (e.g. \cite{CJ89}, \cite{CJ90}, \cite{CR94}, \cite{LANG90}, Figure \ref{Edge}). We can now tentatively interpret these plots as finding an optimal between two compression techniques: non-stochastic models and mixed models. Further research has to show whether these plots can indeed be explained by the theoretical framework developed in this paper. The plots have sharp cut-off points in the area where non-stochastic models collapse into mixed models and vice versa. The point that is in the literature referred to as the edge of chaos (\cite{LANG90}), seems to have no special meaning. It is a collection of ad hoc measurements in an area where the probability decays exponentially. Close to the point $\varphi(x)=|x|/2$ there will be so-called absolutely non-stochastic objects, that exist, but are never observed for strings of any size.

\begin{figure}[!t]
\centering
\includegraphics[width=3in]{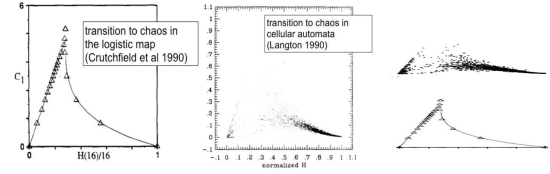}
\caption{So-called 'edge of chaos' phenomena in different domains (With thanks to E. Schultes and the Atlas of Complexity project) On the right the same plots re-scaled to the unitsquare. The analysis in this paper suggests that density of the plots is more relevant than contour lines.}
\label{Some-saw}
\end{figure}

\subsection{Factic processes}

By nature deterministic processes can not generate new information. Information is associated with uncertainty, but in a deterministic process the future is known completely. On the other hand thermodynamic processes seem to increase information in an uncontrollable manner. The amount of information grows but the amount of facticity, or self-descriptive information, is minimal in the end. This suggests that, apart from growth of entropy there is a second useful way to classify processes in terms of the growth of facticity.

Let $x$ be some system that evolves over time and let $x_t$ be the binary description of $x$ system at time $t$: in terms of entropic and factic behavior over time we can distinguish the following five cases:
\begin{enumerate}
  \item Information discarding processes: $\frac{\Delta K(x_t)}{\Delta t} < 0$, $\frac{\Delta \varphi(x_t)}{\Delta t} < 0$
  \item Self-organizing processes: $\frac{\Delta K(x_t)}{\Delta t} < 0$, $\frac{\Delta \varphi(x_t)}{\Delta t} \geq 0$
  \item Reversible processes:$\frac{\Delta K(x_t)}{\Delta t} = 0$, $\frac{\Delta \varphi(x_t)}{\Delta t} = 0$
  \item Random processes: $\frac{\Delta K(x_t)}{\Delta t} > 0$, $\frac{\Delta \varphi(x_t)}{\Delta t} \leq 0$
  \item Factic processes: $\frac{\Delta K(x_t)}{\Delta t} > 0$, $\frac{\Delta \varphi(x_t)}{\Delta t} > 0$
\end{enumerate}
Let's discuss these cases briefly:
\begin{enumerate}
  \item Information discarding processes both decrease entropy and meaningful information and thus violate the second law of thermodynamics and do not occur in closed systems. Standard computing is an example. Recursive functions are in general information discarding functions. When a computation is finished we are left with less entropy and meaningful information as before. Consider adding two numbers: $a + b = c$. Before the computation we have $\log a + \log b$ bits of information, after the computation only $\log (a + b)$.
  \item Self-organizing processes still decrease entropy and do not occur in closed systems. They reduce complexity and increase facticity. The growth of plants in a greenhouse, or bacteria on a petri dish are examples.
  \item Reversible computation in which we keep track of all information (including the meaningful information) is a borderline case. In principle such computations could be energy neutral in a closed system.
  \item Random processes like flipping a coin or diffusion of gases increase entropy but do not generate any new facticity. They are studied in the general theories of randomness and thermodynamics.
  \item Factic processes maintain a balance between model and ad-hoc information, i.e. their data sets can at any time both be interpreted as stochastic and non-stochastic. Processes that both increase entropy and meaningful information have, to my knowledge, not been studied well up till now. Still they deserve our attention: learning, game playing, the development of exchange rates, evolutionary processes and creative processes are factic.
\end{enumerate}

It is clear that factic processes in closed finite domains (e.g. strings of length $n$) cannot increase model information indefinitely (c.f. lemma \ref{Minimality2}). The existence of absolutely non-stochastic uncomputable strings implies that the strings with the largest models will not be found in finite time. In the limit factic processes in finite domains will slowly die out and no increase of model size is possible, no matter how much computing time is spent. Factic processes defy some of our basic philosophical intuitions as the following theorem shows:
\begin{theorem}\label{factic}
In the limit factic processes in infinite domains have no stable model.
\end{theorem}
Proof: facticity is defined as
\[\varphi_U(x)= \min \{|i|: \exists(p)(|\overline{\imath}p|= K_2(x) \ \& \ U(\overline{\imath}p)=x)\}\]
For factic processes we have $\frac{\Delta K(x_t)}{\Delta t} > 0$, $\frac{\Delta \varphi(x_t)}{\Delta t} > 0$. Suppose that at time $t$ an optimal description of $x_t$ is $T_{i,t}(p_{t})=x_{t}$ and that some time $t'$ later we have $T_{i,t'}(p_{t'})=x_{t'}$. Because of increasing entropy and facticity we have $K(i_{t'}) > K(i_t)$ and $K(p_{t'}) > K(p_t)$. For a process with a sufficient statistic the structural part $K(i)$ will stabilize after a certain period of time, pushing all the growth of information to the non-structural part $p$. With factic processes for which the domain can be expanded indefinitely this will never happen. $\Box$\\

This insight runs against central intuitions of scientific methodology. We always feel that no matter how complex a set of phenomena is, if we gather enough data, in the end we will be able to come up with a model that explains them. For factic processes this is per definition not the case. Any model that explains the data now, is sure to fail, at least partly, somewhere in the future.  \footnote{Note that when factic process reach the limit from theorem \ref{max-model} they can have no purely non-stochastic models anymore. There will always be some residual entropy.} In a certain way it is impossible to predict the development of a factic process, but the definition of facticity itself gives us two rules to guide us when predicting the evolution of a factic process:
\begin{itemize}
\item\emph{ Randomness aversion:} A factic process that appears to be random will structure itself (the factor $H^{-1}(s)$.
\item \emph{Model aversion:} A factic process is maximally unstable when it appears to have a regular model (the factor $H(S)=s$).
\end{itemize}

Note that this also implies that factic processes are disruptive: i.e. they are maximally unstable when they appear to have a model.

\section{Discussion}

For approximation of facticity we have theoretically the same limits as were proved in \cite{Power09}: i.e. the facticity is in principle non-approximable in finite time. We may compress the data set, but there is no guarantee that the randomness deficiency of the model is also improved. In practice however, any learning or compression algorithm that allows us to estimate the balance between the data and the model code with reasonable accuracy can be used in real world data. Such algorithms include decision tree induction, nearest neighbor search, neural networks, grammar induction algorithms, standard compression algorithms etc. (\cite{Adriaans2009}, \cite{Adri06}, \cite{cilibrasi-clustering}).  The philosophical question why learning algorithms that in principle are faulty work reasonably well on real life data is still a matter of debate (\cite{Adriaans2009}, \cite{Grunwald:2007:1}).

We may define a resource bounded version of facticity:
\begin{definition}[t-Facticity]\label{t-facticity}
For any time constructable $t$, the $t$-time bounded facticity of $x$ is:
\[
\varphi^{t}_{U}(x)= \min
\left \{
  \begin{array}{cc}
            |i|: &
    \begin{array}{l}
        \exists(p)(|\overline{\imath}p|= K_2(x) \ \& \ U(\overline{\imath}p)=x) \\
        \emph{ in at most $t(|x|)$ steps}
    \end{array}
  \end{array}
\right \}
\]
\end{definition}
Variants of lemma's \ref{Minimality1} and \ref{Model} hold for any time constructable function. Lemma \ref{Minimality2} does not hold for small $t(|x|)$, i.e. we may not have enough computing time to apply the swap function.

In a previous publication (\cite{Adriaans2009}) \emph{normalized facticity} was defined in terms of classical Kolmogorov complexity $C(x)$ and the randomness deficiency of a string: $\delta(x)= |x| - K(x)$ as: \[\varphi(x)=4 \frac{C(x)}{|x|}\times \frac{|x| - C(x)}{|x|}\] Theorem \ref{DEF} shows that the current definition of facticity
covers exactly the same intuitions: simple and complex strings are penalized and maximal facticity is reached (somewhere) in the middle. The present definition lacks the ad-hoc character of the previous one and ties the notion of facticity directly to the complexity of the optimal model that explains the data.

Theorem \ref{DEF} states that $\varphi$ is a good measure for meaningful information: it gives the optimal separation between the structural and ad-hoc part of the data. All structural information is absorbed by the index of the computation. The facticity of a string is the result of two opposing forces: 1) the prefix-free part of the code is penalized by a $2 \log |x| + 1$ factor and should be as short as possible, 2) concatenation of Turing machines reduces complexity so we should store as much information in the prefix as possible. In this sense a string with high normalized facticity has indeed a high tension between structural and ad-hoc information.

Koppel (\cite{Kop87}, \cite{Kop95}) defined the notion of sophistication that, with hindsight, can be seen as an precursor of facticity:
\begin{definition}[Description]
A description of a string $x$ is a pair $(p,d)$ such that $p$ is a self-delimiting total program, and $x$ is an initial segment of $U(p,d)$. The complexity of $x$ then is \[H(x)=min_{p,d}\{|p|+|d|: (p,d)\emph{ is a description of }x \}\]
\end{definition}
\begin{definition}[c-Sophistication]
\[
soph_c(x) = min_p
\left \{
            \begin{array}{cc}
            |p|: & \begin{array}{l}
                            \emph{there is a $d$ such that $(p,d)$} \\
                            \emph{is a description of $x$ and} \\
                             |p|+|d| \leq H(x) + c
                          \end{array}
            \end{array}
\right \}
\]
\end{definition}
Here $c$ is a significance level. It is clear that the idea to use the prefix code of a two part code optimization as a model for a string is already present here. The introduction of a significance level makes sophistication less general than facticity and there is also no guarantee that sophistication is definite in the sense that I have proved above. An idea that is closely related is the proposal for coarse sophistication in \cite{Ant03}:
\begin{definition}[coarse Sophistication]
\[
csoph_c(x) = min_{p,d}
\left \{
\begin{array}{l}
 2|p|+|d| -C(x): U(p,d)=x \\
 \emph{ and $x$ is total.}
\end{array}
\right \}
\]
\end{definition}
Here the term $|p|$ might be interpreted as the sophistication and $|p|+|d| -C(x)$ as a penalty for how far away we are from the minimal program. Clearly this measure is rough and arbitrary compared to facticity, although the factor $2|p|$ might be seen as a discount for the program code, much in the same spirit as the additional prefix code in the facticity. Many of the results for coarse grained sophistication in \cite{Kop95} presumably can be generalized to hold for facticity, but this is a subject for further study.

In an earlier publication (\cite{GL2003}) Gell-Mann and Lloyd developed the notion of the effective complexity of a string in terms of an \emph{ensemble}, or probability distribution, defined on all strings. Foley (\cite{Foley2010}) explores the consequences of this approach in terms of Bayesian inference. Both approaches lead to a view on effective complexity of a string in terms of a balance between its Kolmogorov complexity and its Shannon information in the ensemble. The exact tradeoff between these two notions of information cannot be treated adequately in this setting. Gell-Mann and Lloyd \cite{GL2003} suggest $K[E]+H[E] = K[x]$ as additional constraint, Foley \cite{Foley2010} introduces a temperature parameter $\alpha$ that also has a certain ad hoc character in his proposal for a general prior, based on a tradeoff between Shannon Information and Kolmogorov complexity: \[P_{\alpha} [E] = c 2^{-(K[E]- \alpha H[E])}\] The results in this note show that a definition of such a notion of effective complexity is not only possible within the framework of algorithmic complexity, but also that this approach is much more concise and leads to better insights in to the nature of this phenomenon.

The proposal of self-dissimilarity as a measure for complexity by Wolpert and Mcready (\cite{WM2007}) has interesting parallels with facticity. Simple data sets do not contain enough information to be self-dissimilar. Random data sets are not dissimilar enough to be complex in this sense. The exact relation between self-dissimilarity and facticity is not clear at this moment and the claim in \cite{GL2003} that self-dissimilarity and effective complexity are the same seems premature. My conjecture would be that high facticity is a necessary but not a sufficient condition for high self-dissimilarity. The reason is that we can use facticity to formulate a level of interestingness of data sets (like works of art) that do not seem to have a high level of vertical self-dissimilarity such as the proposal in \cite{WM2007} captures, i.e. the latter description defines a richer notion of complexity.

Below I discuss some possible objections against the theory:

\begin{itemize}
\item \emph{Objection 1) No adequate separation}: Vit\'anyi \cite{Vit06} suggests a connection with the Minimum Description Length (MDL) principle (\cite{RIS78}, \cite{RIS89}). Let $\mathcal{M}$ be the set of prefix-free programs. Using Bayes' law, the optimal computational model under this distribution would be: \[ M_{map}(x) = \arg max_{M \in \mathcal{M}} \frac{m(M) m(x|M)}{m(x)}\] which can be rewritten as: \[= \arg min_{M \in \mathcal{M}} - \log m(M) - \log m(x|M)\] Here $- \log m(M)$ can be interpreted as the length of the optimal \emph{data-code} in Shannon's sense and  $- \log m(x|M)$ as the length of the optimal \emph{data-to-model code}. Using Levin's coding theorem this can be rewritten as:
\begin{equation}\label{MDL} M_{map}(x) = \arg min_{M \in \mathcal{M}} K(M) + K(x|M) \end{equation} This gives optimal \emph{two-part code compression} of $x$. The term $K(M)$ in this expression is of special interest because it seems to capture the notion of \emph{facticity} introduced above.

It is pointed out in \cite{MA2003} that, as soon as we try to separate meaningful information from non-structured information, it is not clear that we can make an objective choice. This is really an issue that has to do with the interplay between one-part and two-part code optimization. We pay a price for the identification of a model:  $K(M) + K(x|M) > K(x)$. Suppose that we want to model strings in the domain $\mathcal{P}(\{0,1\}^k)$ (the set of all sets of binary strings of length $k$). There are two possibilities to formulate an optimal model for a string $x$.
\begin{itemize}
  \item Case 1: If the string is random then the optimal model $M$ would be $\{0,1\}^k$, with $K(M)= O(1)$ and $K(x|M)= |x| + O(1)$, i.e. the index of $x$ in  $\{0,1\}^k$. This gives \[\varphi_{MDL}(x)= |M_{map}(x)| = O(1)\]
  \item Case 2: Equally possible in this context would be a model $M' = \{x\}$, with $K(M)=|x|+O(1)$ and $K(x|M)=O(1).$ This gives: \[\varphi_{MDL}(x)= |M_{map}(x)| \approx |x|\]
\end{itemize}
This shows that the standard MDL formulation does not favor short models. It only favors \emph{optimal separation} between structural and non-structural information. The conclusion is that standard MDL-Kolmogorov theory can not be used as a foundation of the theory of facticity, although in practice it works often quite well. In this paper I present a version, $\varphi$, that is definite, i.e. it gives indeed an objective separation between structural and ad-hoc data.

\item \emph{Objection 2) Pathological Indexes}: This has been called the 'nickname problem' by Gell-Mann. The universal Turing machine that we choose can use pathological index functions to select the specific Turing machines. Specifically there are choices that give an index of length $1$ to a universal Turing machine $U$. In this case we get a new universal model for one bit. Our MDL code would always select this universal model coded in $1$ bit and continue with standard one-part Kolmogorov code for $U$. Clearly we have to put constraints on the definition of the indexes. An ideal index function for measuring facticity would have to observe two seemingly conflicting conditions:
    \begin{itemize}
      \item It should reflect the Kolmogorov complexity of the definition of the underlying Turing machine and
      \item be computable
    \end{itemize}
I have shown that such an index cannot exist, but that 'faithfulness' is a reasonable assumption in most cases.

\item \emph{Objection 3) The models are not cognitively relevant} (\cite{MA2003}). Since Kolmogorov complexity gives the complexity of individual objects there is no guarantee that the part of the description we single out as the model, captures aspects that from a more cognitively relevant point of view would be seen as model information. If we develop an optimal two-part code description of an individual horse with three legs, the fact that this animal has three legs might well end up in the 'model' part of the description, although the ideal horse still would have four legs. This is true, but when we only have data about a horse with three legs, then this is how it should be. Facticity captures the notion of an optimal model from an algorithmic point of view. Whether these models are cognitively relevant is subject for further study, but given their generality one may expect this (\cite{Wolff06}, \cite{Chat03}).

\item \emph{Objection 4) Choice for $U$ introduces a bias.} This is true but the length of the theories generated by different choices of Turing machines always will at maximum only be a constant apart (\cite{LiVi08}). So asymptotically the different measures are still comparable. Suppose there exists a universal Turing machine $T_i$ with, an index of length $i$, that would generate a considerable smaller code $p'$ explaining a certain string $x$ than our current choice $T_j$ with $p$. We simply compress the program by prefixing $p'$ with the self-delimiting code for the index $i$ and feed this in to $T_j$. The proof of theorem \ref{DEF} shows that given a choice for a reference machine $U$ facticity is defined with exact precision.

\item \emph{Objection 5) There are different models with equal facticity.} There may be several competing models that compress the data equally good. This is actually a feature more than a bug. Gestalt switches are a concrete example where two incompatible models give an equally good interpretation of the data. So we do not want a theory that restricts itself to solutions where only one model is the best.

\item \emph{Objection 6) No physical substrate.} "The effective complexity of a string as a purely formal construct, lacking a physical interpretation, is either close to zero, or equal to the string's algorithmic complexity, or arbitrary, depending on the auxiliary criterion chosen to pick out the regular component of the string\cite{MA2003}." This observation is simply wrong, as the central result in this paper proves.
\end{itemize}

\section{Conclusion}
In this paper it is shown that it is possible and promising to develop a theory of meaningful self-information of strings based on Kolmogorov complexity. Such a theory allows us to define a concept (of course there are many others possible and useful) of meaningful information in terms of facticity. Further research will involve an analysis of processes that create data sets with high facticity (games, genetic algorithms, the stock market, evolution) analysis of existing data sets in the light of this theory \cite{CJ89}, \cite{CJ90}, \cite{CR94}, \cite{LANG90}) and the development of a theory of conditional facticity.


%



\section*{Acknowledgment}

The author would like to thank Duncan Fowley, David Oliver, Amos Golan, Ariel Caticha, Peter van Emde Boas, Erik Schultes, Harry Buhrman, Luis Antunes, Andr\'e Souto and Peter Bloem for discussions on the ideas presented in this paper. This research was partly supported by the Dutch Virtual Lab for e-Science (VL-e) project, by the Dutch national program COMMIT, a Templeton Foundation's Science and Significance of Complexity Grant supporting The Atlas of Complexity Project and the Info-Metrics Institute of the American University in Washington.

\ifCLASSOPTIONcaptionsoff
  \newpage
\fi



%

%


\begin{IEEEbiographynophoto}{Pieter Adriaans} (1955) studied philosophy in Leiden. From 1989 till 2001 he was research director at Syllogic, a company specializing in AI, data mining and learning systems. In  1992 he obtained his PhD  from the University of Amsterdam on a study on grammar learning under supervision of Peter Van Emde Boas.  In 1997 Syllogic was sold to Perot Systems. Since 1997 Adriaans is professor of learning and adaptive systems at the university of Amsterdam. He is editor of the Handbook of Philosophy of Information (Elsevier 2008). His main interests are complexity theory, learning theory and philosophy of information.
\end{IEEEbiographynophoto}






\end{document}